\documentstyle[a4,12pt]{article}
\begin{document}
\parskip=5pt plus 1pt minus 1pt

\begin{flushright}
{\bf DPNU-98-12} \\
{April 1998}
\end{flushright}

\vspace{0.2cm}
\begin{center}
{\Large\bf A Phenomenological Interpretation of 
Atmospheric \\ and Solar Neutrino Oscillations} 
\footnote{Talk given at the Workshop on Fermion Masses
and $CP$ Violation, Hiroshima, March 5 - 6, 1998}
\end{center}

\vspace{0.4cm}
\begin{center}
{\bf Zhi-zhong Xing} 
\footnote{ Electronic address: xing@hep.physik.uni-muenchen.de} \\
{\sl Sektion Physik, Universit$\sl\ddot{a}$t M$\sl\ddot{u}$nchen,
Theresienstrasse 37, 80333 M$\sl\ddot{u}$nchen, Germany} \\
{\sl and ~ Department of Physics, Nagoya University, Chikusa--ku, Nagoya
464--01, Japan}
\end{center}

\vspace{1.7cm}
\begin{abstract}
We give a phenomenological interpretation of atmospheric and solar
neutrino oscillations by considering two scenarios for the neutrino
mass spectrum: (a) $m_1 \ll m_2 \ll m_3$ and (b) $m_1 \approx m_2 \approx
m_3$. A new parametrization of the flavor mixing matrix, which can
naturally reflect the hierarchy of lepton masses in scenario (a) and 
the approximate decoupling of solar and atmospheric neutrino
oscillations, is highlighted. For scenario (b) an ansatz starting with flavor
democracy for charged leptons and mass degeneracy for neutrinos is
proposed, and two different symmetry breaking possibilities with
``maximal calculability'' are discussed. Finally we point out that
possible $(\sin^2 2\theta, \Delta m^2)$ parameter-space correlation should
seriously be taken into account in future analyses of neutrino oscillations. 
\end{abstract}

\newpage

\section{Introduction}
\setcounter{equation}{0}

The recent preliminary results of the Superkamiokande experiment
\cite{SuperK} have
provided a more credible hint that the atmospheric neutrino anomaly
should be attributed to the oscillation of massive neutrinos. The
possible oscillation scenario could {\it a priori} be either $\nu_\mu
\leftrightarrow \nu_e$ or $\nu_\mu \leftrightarrow \nu_\tau$, but the 
former appears disfavored by the result of the CHOOZ experiment \cite{CHOOZ}. It
turns out that the most likely solution to the atmospheric neutrino
problem comes from the $\nu_\mu \leftrightarrow \nu_\tau$ oscillation
with the following ranges of two oscillation parameters \cite{SuperK,Atm}:
\begin{eqnarray}
&& \Delta m^2_{\rm atm} \; \approx \; ( 0.3 ~ - ~ 8 ) \times 10^{-3} ~ {\rm
eV}^2 \; , \nonumber \\
&& \sin^2 2 \theta_{\rm atm} \; \approx \; 0.7 ~ - ~ 1 \; .
\end{eqnarray}
The large mixing angle here implies that the physics responsible for
neutrino masses and lepton flavor mixings may be essentially
different from that for the quark sector.

The long-standing problem of the solar neutrino deficit has been
evidenced again by the preliminary data from Superkamiokande \cite{SuperK}. In the
assumption of the resonant Mikheyev-Smirnov-Wolfenstein (MSW) effect
\cite{MSW}, the recent analysis of all
available solar neutrino data gives the ranges of two oscillation
parameters as follows \cite{Solar}
\footnote{The large-angle MSW solution \cite{Solar}, which allows $\Delta m^2_{\rm
sun} \approx (0.8 ~ - ~ 3.0) \times 10^{-5} ~ {\rm eV}^2$ and $\sin^2
2\theta_{\rm sun} \approx 0.42 ~ - ~ 0.74$, will not be considered in
this work. Also we shall not consider the result from the LSND experiment \cite{LSND},
since a complete interpretation of all existing neutrino
oscillation data requires the introduction of a sterile neutrino.}:
\begin{eqnarray}
&& \Delta m^2_{\rm sun} \; \approx \; ( 0.3 ~ - ~ 1.2 ) \times 10^{-5} ~ {\rm
eV}^2 \; , \nonumber \\
&& \sin^2 2 \theta_{\rm sun} \; \approx \; (0.3 ~ - ~ 1.2) \times 10^{-2} \; .
\end{eqnarray}
It is also possible to interpret the solar neutrino deficit by
invoking the pure vacuum oscillation mechanism (the ``Just-so''
solution). In this case the oscillation parameters read \cite{Solar}:
\begin{eqnarray}
&& \Delta m^2_{\rm sun} \; \approx \; ( 0.6 ~ - ~ 1.1 ) \times 10^{-10} ~ {\rm
eV}^2 \; , \nonumber \\
&& \sin^2 2 \theta_{\rm sun} \; \approx \; 0.7 ~ - ~ 1 \; ,
\end{eqnarray}
for the survival probability of electron neutrinos.

Note that in current analyses $\Delta m^2$ and $\sin^2 2\theta$ are treated as two 
independent parameters. This might be misleading in some models of lepton mass generation,
where the mixing angle $\theta$ depends sensitively on the neutrino mass ratios --
the oscillation parameters $\sin^2 2\theta$ and $\Delta m^2$ are correlated with each
other.

A deeper insight into the yet unknown dynamics of the lepton mass
generation requires nontrivial steps beyond the standard
model. Phenomenologically the proper approach might be first to
identify the patterns of lepton mass matrices from some kinds of
symmetries, and then to interpret the neutrino oscillation data. For
this purpose, two distinct possibilities of the neutrino mass spectrum
are of particular interest and have attracted some attention \cite{Review}: 

(a) Three neutrino masses perform a clear hierarchy: $m_1 \ll m_2
\ll m_3$ ; 

(b) Three neutrino masses are almost degenerate: $m_1 \approx m_2
\approx m_3$ . \\
In comparison with the well-known mass spectrum of the charged
leptons ($m_e \ll m_\mu \ll m_\tau$), we expect that the 
charged lepton and neutrino mass matrices ($M_l$ and $M_\nu$)  
in scenario (a) may both take the
hierarchical form, similar to the quark mass matrices. The neutrino
mass matrix in scenario (b), however, must take a form different from the
charged lepton mass matrix.

Phenomenologically a popular approach to describing hierarchical
neutrino masses might be to start from the Fritzsch-like ans$\rm\ddot{a}$tze 
\cite{F78,Fritzsch} with or without the see-saw mechanism \cite{See-Saw}. In
the assumption of the charged lepton flavor democracy and the neutrino 
mass degeneracy, a model has also been proposed to accommodate the almost 
degenerate neutrinos \cite{FX96,FTY}. 
In this talk we shall have a further look at these two
neutrino mass scenarios and to confront them with the updated
results of neutrino oscillations. We highlight a new parametrization of
the $3\times 3$ flavor mixing matrix for scenario (a), and find that its
three mixing angles can get apparent (physical) meanings from the
Fritzsch-like ans$\rm\ddot{a}$tze as well as from the approximately decoupled
atmospheric and solar neutrino oscillations. For scenario (b) we first point out 
a few more possibilities to break the mass degeneracy of neutrinos
which respect the ``maximal calculability'' requirement but were not
considered in Refs. \cite{FX96,FTY}, and then discuss their
consequences on neutrino oscillations. Finally we comment briefly on
the possible problem of $(\sin^2 2\theta, \Delta m^2)$ parameter-space 
correlation in current analyses of neutrino oscillations.

\section{Hierarchical neutrino masses}
\setcounter{equation}{0}

An analogy between the lepton mass hierarchy and the quark
mass hierarchy is attractive in phenomenology, and this could naturally
be obtained from a grand unified theory responsible for the generation 
of both quarks and leptons. It has been noticed, among a variety 
of parametrizations of the quark flavor mixing matrix, that there is a
particular parametrization which can naturally reflect the hierarchy
of quark masses and is favored by the $B$-meson physics \cite{FX97}:
\begin{eqnarray}
V & = & \left ( \matrix{
c_{\rm u}	& s_{\rm u}	& 0 \cr
-s_{\rm u}	& c_{\rm u}	& 0 \cr
0	& 0	& 1 \cr } \right )  \left ( \matrix{
e^{-{\rm i}\varphi}	& 0	& 0 \cr
0	& c	& s \cr
0	& -s	& c \cr } \right )  \left ( \matrix{
c_{\rm d}	& -s_{\rm d}	& 0 \cr
s_{\rm d}	& c_{\rm d}	& 0 \cr
0	& 0	& 1 \cr } \right )  \nonumber \\ \nonumber \\
& = & \left ( \matrix{
s_{\rm u} s_{\rm d} c + c_{\rm u} c_{\rm d} e^{-{\rm i}\varphi} &
s_{\rm u} c_{\rm d} c - c_{\rm u} s_{\rm d} e^{-{\rm i}\varphi} &
s_{\rm u} s \cr
c_{\rm u} s_{\rm d} c - s_{\rm u} c_{\rm d} e^{-{\rm i}\varphi} &
c_{\rm u} c_{\rm d} c + s_{\rm u} s_{\rm d} e^{-{\rm i}\varphi}   &
c_{\rm u} s \cr
- s_{\rm d} s	& - c_{\rm d} s	& c \cr } \right ) \; ,
\end{eqnarray}
where $s_{\rm u} \equiv \sin\theta_{\rm u}$, $s_{\rm d} \equiv \sin\theta_{\rm d}$,
$c \equiv \cos\theta$, etc. One can see 
\begin{eqnarray}
\tan\theta_{\rm u} & = & \left | \frac{V_{ub}}{V_{cb}} \right | \; ,
\nonumber \\
\tan\theta_{\rm d} & = & \left | \frac{V_{td}}{V_{ts}} \right | \; ,
\end{eqnarray}
and
\begin{equation}
\sin\theta \; =\; \sqrt{|V_{ub}|^2 ~ + ~ |V_{cb}|^2} \; ;
\end{equation}
i.e., all three mixing angles can be directly measured from $B$ decay
or $B^0$-$\bar{B}^0$ mixing experiments. An analysis of current data
on flavor mixing and $CP$ violation gives $\theta = 2.30^{\circ} \pm
0.09^{\circ}$, $\theta_{\rm u} = 4.87^{\circ} \pm 0.86^{\circ}$,
$\theta_{\rm d} = 11.71^{\circ} \pm 1.09^{\circ}$, and $\varphi =
91.1^{\circ} \pm 11.8^{\circ}$ \cite{Stocchi}. For a variety of quark
mass matrices, each of the four parameters gets a definite (physical)
meaning \cite{FX97,FX98}: $\varphi$ amounts approximately to the phase difference
between the up and down mass matrices, $\theta$ essentially describes the heavy
quark flavor mixing, and $\theta_{\rm u}$ and $\theta_{\rm d}$ are
related to the light quark mass ratios to a good degree of accuracy \cite{FX98}:
\begin{eqnarray}
\tan \theta_{\rm u} & = & \sqrt{\frac{m_u}{m_c}} \;\; , \nonumber \\
\tan \theta_{\rm d} & = & \sqrt{\frac{m_d}{m_s}} \;\; .
\end{eqnarray}
These simple results make the parametrization (2.1) uniquely useful
for the study of $B$-meson physics and quark mass matrices. In
contrast, we argue that the parametrization advocated by
the Particle Data Group \cite{PDG96} indeed has little merit to be
``standard''.

In view of our wealthy knowledge on quark mass
matrices and quark flavor mixings, we believe that the most appropriate
description of the $3\times 3$ lepton flavor mixing matrix, in terms
of three Euler angles ($\theta_l$, $\theta_\nu$, $\theta$) and one
$CP$-violating phase $\phi$, should read as follows \cite{FX99}:
\begin{eqnarray}
V & = & \left ( \matrix{
c^{~}_l	& s^{~}_l	& 0 \cr
-s^{~}_l	& c^{~}_l	& 0 \cr
0	& 0	& 1 \cr } \right )  \left ( \matrix{
e^{-i\phi}	& 0	& 0 \cr
0	& c	& s \cr
0	& -s	& c \cr } \right )  \left ( \matrix{
c_{\nu}	& -s_{\nu}	& 0 \cr
s_{\nu}	& c_{\nu}	& 0 \cr
0	& 0	& 1 \cr } \right )  \nonumber \\ \nonumber \\
& = & \left ( \matrix{
s^{~}_l s_{\nu} c + c^{~}_l c_{\nu} e^{-i\phi} &
s^{~}_l c_{\nu} c - c^{~}_l s_{\nu} e^{-i\phi} &
s^{~}_l s \cr
c^{~}_l s_{\nu} c - s^{~}_l c_{\nu} e^{-i\phi} &
c^{~}_l c_{\nu} c + s^{~}_l s_{\nu} e^{-i\phi}   &
c^{~}_l s \cr - s_{\nu} s	& - c^{~}_l s	& c \cr } \right ) \; ,
\end{eqnarray}
where $s^{~}_l \equiv \sin\theta_l$, $s_{\nu} \equiv \sin\theta_{\nu}$, 
$c \equiv \cos\theta$, etc. This parametrization has naturally
reflected the hierarchical properties of lepton masses, and its
mixing angles may also have very instructive (physical) meanings. To see this point 
more clearly, we consider two reasonable limits for the sake of the
assumed neutrino mass hierarchy.

(1) The ``light lepton'' limit: $m_e/m_\mu \rightarrow 0$ and $m_1/m_2 \rightarrow
0$. In this case both the charged lepton mass matrix $M_l$ and the
neutrino mass matrix $M_\nu$ have texture zeros in the
$(1,i)$ and $(i,1)$ positions for $i=1,2$ and $3$. Therefore the
mixing angles $\theta_l$ and $\theta_\nu$ vanish, and $V$ exclusively
describes the mixing between the 2nd and 3rd lepton families. The
resultant probability for $\nu_\mu \rightarrow \nu_\tau$ transition reads
\begin{equation}
P (\nu_\mu \rightarrow \nu_\tau) \; =\; \sin^2 2\theta ~ \sin^2 \left
( 1.27 ~ \frac{\Delta m^2_{23} L}{|{\bf P}|} \right ) \; \; ,
\end{equation}
where $\Delta m^2_{23} \equiv m^2_3 - m^2_2$ (in unit ${\rm eV}^2$), $L$ 
is the distance from the production point of the muon neutrino to its 
interaction point (in unit km), and $\bf P$ is the momentum of the neutrino beam
(in unit GeV). The transition probabilities for $\nu_e \rightarrow
\nu_\mu$ and $\nu_e \rightarrow \nu_\tau$ both vanish in this
``light lepton'' limit. One can see that $\theta = \theta_{\rm atm}$ and
$\Delta m^2_{23} = \Delta m^2_{\rm atm}$, if the atmospheric neutrino
anomaly is ascribed to $\nu_\mu \rightarrow \nu_\tau$.

(2) The ``heavy lepton'' limit: $m_\mu/m_\tau \rightarrow 0$ and $m_2/m_3
\rightarrow 0$. In this case the $(\tau, \nu_\tau)$ system is strongly 
decoupled from the light $(\mu, \nu_\mu)$ and $(e, \nu_e)$ systems due 
to the dominance of the 3rd family in the lepton mass spectrum. The
transition probabilities for $\nu_\mu \rightarrow \nu_\tau$ and $\nu_e 
\rightarrow \nu_\tau$ vanish, while the survival probability $P(\nu_e
\rightarrow \nu_e)$ is given as
\begin{equation}
P (\nu_e \rightarrow \nu_e) \; =\; 1 ~ - ~ \sin^2 2\theta^{~}_{\rm C} ~ \sin^2 \left
( 1.27 ~ \frac{\Delta m^2_{12} L}{|{\bf P}|} \right ) \; \; ,
\end{equation}
where $\Delta m^2_{12} \equiv m^2_2 - m^2_1$, and the ``leptonic''
Cabibbo angle $\theta_{\rm C}$ reads
\begin{equation}
\sin\theta_{\rm C} \; =\; \sqrt{s^2_l c^2_\nu + c^2_l s^2_\nu - 2 s^{~}_l
c^{~}_l s_\nu c_\nu \cos \phi} \;\; .
\end{equation}
Considering the mass hierarchy ($m_e/m_\mu \ll 1$ and $m_1/m_2 \ll
1$), we expect $\theta_{\rm C}$ to be small enough to match the small-angle
MSW solution to the solar neutrino deficit. Then $\theta_{\rm C} =
\theta_{\rm sun}$ and $\Delta m^2_{12} = \Delta m^2_{\rm sun}$ hold.

In both limits taken above, $CP$ violation vanishes. To examine how
far these limits are from the reality, we have to assume the
explicit pattern of $M_l$ and $M_\nu$ and to express the mixing angles 
$(\theta_l, \theta_\nu, \theta)$ in terms of the lepton mass ratios.
Only if we impose the phenomenological constraints 
$\Delta m^2_{23} = \Delta m^2_{\rm atm}$ and $\Delta m^2_{12} = \Delta
m^2_{\rm sun}$, however, we can always obtain $m_3 \approx (1.7 ~ - ~ 9) \times
10^{-2}$ eV from (1.1) and $m_2 \approx (1.7 ~ - ~ 3.5) \times 10^{-3}$ eV
from (1.2) because of the assumed hierarchy $m_1 \ll m_2 \ll m_3$.  
The large gap between $\Delta m^2_{\rm atm}$ and $\Delta m^2_{\rm
sun}$ generally allows the oscillations of solar and atmospheric
neutrinos to decouple as a good approximation, provided the flavor mixing 
matrix element $V_{e3} = s^{~}_l s$ is sufficiently small \cite{Giunti}. As we
can see from some reasonable Fritzsch-like ans$\rm\ddot{a}$tze of
lepton mass matrices \cite{F78,Fritzsch}, the
mixing angle $\theta_l$ is essentially given by \cite{FX99}
\begin{equation}
\theta_l \; =\; \arctan \left ( \sqrt{\frac{m_e}{m_\mu}} \right ) \; \approx \;
4^{\circ} \; .
\end{equation}
Therefore the decoupling
condition for the solar and atmospheric neutrino oscillations can be
satisfied, and the instructive formulas obtained in the above two
limits should hold to a good degree of accuracy.

In this talk we shall not go into the details of the neutrino mass
matrix with hierarchical mass eigenvalues (for a recent study with
extensive references, see, e.g., Ref. \cite{FX99}).

\section{Nearly degenerate neutrino masses}
\setcounter{equation}{0}

The dominance of $m_\tau$ in the hierarchical (``H'') mass spectrum of 
charged leptons implies a plausible limit in which the mass matrix
takes the form
\begin{equation}
M^{\rm H}_l \; =\; c^{~}_l \left (\matrix{
0	& 0	& 0 \cr
0	& 0	& 0 \cr
0	& 0	& 1 \cr} \right ) \; 
\end{equation}
with $c^{~}_l = m_\tau$. This mass matrix is equivalent to the
following matrix with $S(3)_{\rm L}\times S(3)_{\rm R}$ symmetry or
flavor democracy (``D''):
\begin{equation}
M^{\rm D}_l \; =\; \frac{c^{~}_l}{3} \left (\matrix{
1	& 1	& 1 \cr
1	& 1	& 1 \cr
1	& 1	& 1 \cr} \right ) \; 
\end{equation}
through the orthogonal transformation $O^{\dagger} M^{\rm H}_l O =
M^{\rm D}_l$, where
\begin{equation}
O \; =\; \left (\matrix{
\displaystyle \frac{1}{\sqrt{2}}	& - \displaystyle \frac{1}{\sqrt{2}}	& 0 \cr\cr
\displaystyle \frac{1}{\sqrt{6}}	& \displaystyle \frac{1}{\sqrt{6}}	
& - \displaystyle \frac{2}{\sqrt{6}} \cr\cr
\displaystyle \frac{1}{\sqrt{3}}	& \displaystyle
\frac{1}{\sqrt{3}}	& \displaystyle \frac{1}{\sqrt{3}}
\cr} \right ) \; .
\end{equation}
For either $M^{\rm H}_l$ or $M^{\rm D}_l$, further symmetry breaking
terms can be introduced to generate masses for muon and
electron. Recently the possible significance of the approximate
democratic mass matrices has been emphasized towards understanding
fermion masses and flavor mixings \cite{Democracy}.

The similar picture is however invalid for the neutrino sector, if
three neutrino masses are almost degenerate. Provided the netrino
masses persist in an exact degeneracy (``D'') symmetry, the corresponding mass
matrix should take the diagonal form: 
\begin{equation}
M^{\rm D}_\nu \; =\; c_\nu \left (\matrix{
\eta_1	& 0	& 0 \cr
0	& \eta_2	& 0 \cr
0	& 0	& \eta_3 \cr} \right ) \; ,
\end{equation}
where $c_\nu \equiv m_0$ measures the mass scale of three
neutrinos, and $\eta_i$ ($=\pm 1$) denotes the $CP$-parity of the Majorana
neutrino $\nu_i$.  
By breaking the mass degeneracy of $M^{\rm D}_\nu$ slightly, one may get the realistic
neutrino masses $m_1$, $m_2$ and $m_3$ (at least two of them are
different from $m_0$ and different from each other).

A purely phenomenological assumption, that a more fundamental theory
of lepton interactions might simultaneously accommodate the charged
lepton mass matrix $M^{\rm D}_l$ and the neutrino mass matrix $M^{\rm D}_\nu$
in the symmetry limit, has been made first in Ref. \cite{FX96} and
then in Ref. \cite{FTY} to discuss
the lepton flavor mixing and to interpret the neutrino oscillation
data. In this case one can find that the constant matrix $O$ may play
an important role in the flavor mixing matrix, obtained fom the diagonalization
of $M^{\rm D}_l$ and $M^{\rm D}_\nu$. 
The realistic flavor mixing matrix $V$ 
depends on the explicit pattern of the perturbative corrections to
$M^{\rm D}_l$ and $M^{\rm D}_\nu$. 
For simplicity we shall neglect the small correction from
nonvanishing $m_e/m_\mu \ll 1$ and $m_\mu/m_\tau \ll 1$, and concentrate on the
degeneracy-breaking pattern of $M_\nu$.

The forms of the perturbative correction to $M^{\rm D}_\nu$, denoted as $\Delta
M_\nu$, can be classified into two categories.

(i) $\Delta M_\nu$ is a completely diagonal matrix with two or three
parameters. In this case $M^{\rm D}_\nu + \Delta M_\nu$ remains diagonal,
then the flavor mixing
matrix $V$ turns out to be $O$. As shown in Ref. \cite{FX96}, this
ansatz results in the survival probability of the electron neutrino as
\begin{equation}
P (\nu_e \rightarrow \nu_e ) \; \approx \; 1 ~ - ~ \sin^2 \left (1.27
~ \frac{\Delta m^2_{12} L}{|{\bf P}|} \right ) \; ;
\end{equation}
i.e., the mixing factor $\sin^2 2 \theta_{\rm sun}$ is almost
maximal. Obvioulsy this result is favored by the vacuum oscillation
solution to the solar neutrino deficit, which requires the
mass-squared difference $|\Delta m^2_{12}| = \Delta m^2_{\rm sun} \sim 
10^{-10} ~ {\rm eV}^2$ (see (1.3) for the value). 
The transition probability of $\nu_\mu \rightarrow \nu_\tau$, on the
other hand, reads as follows:
\begin{equation}
P (\nu_\mu \rightarrow \nu_\tau ) \; \approx \; \frac{8}{9} ~ \sin^2 \left (1.27
~ \frac{\Delta m^2_{23} L}{|{\bf P}|} \right ) \; ,
\end{equation}
where the oscillating term of $\Delta m^2_{12}$ has been ignored due
to its tiny effect, and $\Delta m^2_{23} \approx \Delta m^2_{13}$ is a 
good approximation because of $\Delta m^2_{\rm atm} \gg |\Delta
m^2_{12}|$. We conclude that both the atmospheric and solar neutrino
oscillations can be interpreted with this lepton mass ansatz, which
allows a remarkable mass degeneracy between $v_1$ and $v_2$ states as
well as a relatively weaker mass degeneracy between $v_2$ and $\nu_3$
states.

For Majorana neutrino masses, it is necessary to fulfill the bound $\langle 
m \rangle \leq 0.7$ eV for neutrinoless $\beta\beta$-decay \cite{Beta}, where 
$\langle m \rangle$ is an effective mass factor. It is known that the 
$(\beta\beta)_{0\nu}$-decay amplitude depends on the masses of Majorana neutrinos 
$m_{i}$ and on the elements of the lepton mixing matrix $V_{ei}$:
\begin{equation}
\langle m \rangle \; \sim \; 
\sum^{3}_{i=1}\left ( V_{ei}^{2} ~ m_i ~ \eta_i \right ) \; ,
\end{equation}
where $\eta_i$ is the $CP$-parity of the Majorana field for $\nu_i$. 
If $\eta_1 =+1$ and $\eta_2 =-1$ (or vice versa), one finds 
that $\langle m \rangle \sim (m_{1}-m_{2})/2$, which is considerably suppressed 
due to the near degeneracy of $m_{1}$ and $m_{2}$. This implies that there may 
exist two Majorana neutrinos with opposite $CP$ eigenvalues, and their relative 
$CP$ parities are in principle observable in $(\beta\beta)_{0\nu}$-decay.
Within our approach this possibility exists. Thus a degenerate Majorana mass of 
about $m_0 \sim 2.5$ eV for all three neutrinos, favored by the hot
dark matter, need not be in conflict with the data on neutrinoless $\beta\beta$-decay.

(ii) $\Delta M_\nu$ is not a completely diagonal matrix but consists of
at least two parameters. In this case a variety of patterns for
$\Delta M_\nu$ is allowed. To ensure the ``maximal calculability'' for the 
neutrino mass matrix, we require a special form of $\Delta M_\nu$
which has two unknown parameters (to break the mass degeneracy of
$M^{\rm D}_\nu$) and can be diagonalized by a {\it constant} orthogonal
transformation (independent of the neutrino masses). Then we find that
only three patterns of $\Delta M_\nu$, as listed in Table 1, 
satisfy these strong requirements.
They can be diagonalized by three Euler rotation matrices $R_{ij}$
with the rotation angles $\theta_{ij} = 45^{\circ}$ in the (1,2),
(2,3) and (3,1) planes respectively (see also Table 1). 
In Ref. \cite{FTY} only pattern (I) was proposed and discussed. For each pattern 
the resultant flavor mixing matrix reads as
$V \approx O R_{ij}$,
which remains a constant matrix before the introduction of small
corrections from $m_e/m_\mu$ and $m_\mu/m_\tau$. We calculate the transition
probability for $\nu_\mu \rightarrow \nu_\tau$ and find the mixing factor to 
be $8/9$, $2/9$ or $2/9$, respectively, for pattern (I), (II) or (III).
Therefore only pattern (I) can survive when confronting the
atmospheric neutrino data. The survival probability $P(\nu_e
\rightarrow \nu_e)$ amounts to unity in the limit $m_e = m_\mu =0$ for 
pattern (I). It can interpret the small-angle MSW solution to the
solar neutrino deficit, however, after a proper perturbative term $\Delta M_l$ is
introduced to $M^{\rm D}_l$. For example, $\sin^2 2\theta_{\rm sun}
\sim m_e/m_\mu$ may be obtained if $\Delta M_l$ is taken to be of the
diagonal form \cite{FX96,FTY}. Also this ansatz has no conflict with
requirements of neutrinoless $\beta\beta$-decay and hot dark matter
on neutrino masses.
\small
\begin{table}[t]
\caption{Three patterns of $\Delta M_\nu$ and their consequences on
$V$ and $P(\nu_\mu \rightarrow \nu_\tau)$}
\vspace{-0.6cm}
\begin{center}
\begin{tabular}{ccccc} \\ \hline\hline 
Pattern     & $\Delta M_\nu$	& $R_{ij}$	& $V$	& $P(\nu_\mu
\rightarrow \nu_\tau)$ \\  \hline \\
(I)	&
$\left ( \matrix{
0	& \epsilon	& 0 \cr
\epsilon	& 0	& 0 \cr
0	& 0	& \delta } \right ) $	&
$\left ( \matrix{
\displaystyle \frac{1}{\sqrt{2}}	& \displaystyle \frac{1}{\sqrt{2}}	& 0 \cr\cr
- \displaystyle \frac{1}{\sqrt{2}}	& \displaystyle \frac{1}{\sqrt{2}}	& 0 \cr\cr
0	& 0	& 1 \cr } \right ) $	&
$\left ( \matrix{
1	& 0	& 0 \cr\cr
0	& \displaystyle \frac{1}{\sqrt{3}}	& - \displaystyle \frac{2}{\sqrt{6}} \cr\cr
0	& \displaystyle \frac{2}{\sqrt{6}}	
& \displaystyle \frac{1}{\sqrt{3}} \cr } \right ) $	&
$\displaystyle \frac{8}{9} ~ \sin^2 \left ( 1.27 ~ \frac{\Delta m^2_{23} L}{|{\bf
P}|} \right ) $ \\  \\ \hline \\
(II)	&
$\left ( \matrix{
\delta	& 0	& 0 \cr
0	& 0	& \epsilon \cr
0	& \epsilon	& 0 } \right ) $	&
$\left ( \matrix{
1	& 0	& 0 \cr\cr
0	& \displaystyle \frac{1}{\sqrt{2}}	& \displaystyle \frac{1}{\sqrt{2}} \cr\cr
0	& - \displaystyle \frac{1}{\sqrt{2}}	& \displaystyle 
\frac{1}{\sqrt{2}} \cr } \right ) $	&
$\left ( \matrix{
\displaystyle \frac{1}{\sqrt{2}}	& - \displaystyle \frac{1}{2}		
& - \displaystyle \frac{1}{2} \cr\cr
\displaystyle \frac{1}{\sqrt{6}}	& \displaystyle \frac{3}{2 \sqrt{3}}	
& - \displaystyle \frac{1}{2 \sqrt{3}} \cr\cr
\displaystyle \frac{1}{\sqrt{3}}	& 0	
& \displaystyle \frac{2}{\sqrt{6}} \cr } \right ) $	&
$\displaystyle \frac{2}{9} ~ \sin^2 \left ( 1.27 ~ \frac{\Delta m^2_{23} L}{|{\bf
P}|} \right ) $ \\  \\ \hline \\
(III)	&
$\left ( \matrix{
0	& 0	& \epsilon \cr
0	& \delta	& 0 \cr
\epsilon	& 0	& 0 } \right ) $	&
$\left ( \matrix{
\displaystyle \frac{1}{\sqrt{2}}	& 0	& \displaystyle \frac{1}{\sqrt{2}} \cr\cr
0	& 1	& 0 \cr\cr
- \displaystyle \frac{1}{\sqrt{2}}	& 0	& \displaystyle 
\frac{1}{\sqrt{2}} \cr } \right ) $	&
$\left ( \matrix{
\displaystyle \frac{1}{2}	& - \displaystyle \frac{1}{\sqrt{2}}		
& - \displaystyle \frac{1}{2} \cr\cr
- \displaystyle \frac{1}{2 \sqrt{3}}	& \displaystyle \frac{1}{\sqrt{6}}	
& - \displaystyle \frac{3}{2 \sqrt{3}} \cr\cr
\displaystyle \frac{2}{\sqrt{6}}	& \displaystyle
\frac{1}{\sqrt{3}} & 0 \cr } \right ) $	&
$\displaystyle \frac{2}{9} ~ \sin^2 \left ( 1.27 ~ \frac{\Delta m^2_{23} L}{|{\bf
P}|} \right ) $ \\  \\ 
\hline\hline
\end{tabular}
\end{center}
\end{table}
\normalsize

Note that the situation will change if one assume a theory of lepton
interactions to accommodate the charged lepton mass $M^{\rm H}_l$
(instead of $M^{\rm D}_l$) and the neutrino mass matrix $M^{\rm D}_\nu$
simultaneously in the symmetry limit. In this case the large mixing
angle(s) of $V$ can only come from diagonalizing $M^{\rm D}_\nu + \Delta
M_\nu$, where $\Delta M_\nu$ may take the forms listed in Table 1 to
guarantee the ``maximal calculability''. It is obvious that $V \approx
R_{ij}$ holds. We find that only pattern (II) of $\Delta M_\nu$ is
favored by the atmospheric neutrino data; i.e.,
\begin{equation}
V \; \approx \; \left ( \matrix{
1	& 0	& 0 \cr
0	& \displaystyle \frac{1}{\sqrt{2}}	& \displaystyle
\frac{1}{\sqrt{2}} \cr
0	& - \displaystyle \frac{1}{\sqrt{2}}	& \displaystyle
\frac{1}{\sqrt{2}} \cr } \right ) \; 
\end{equation}
in the neglect of the $\Delta M_l$ effect on $M^{\rm H}_l$, and the
corresponding probability of $\nu_\mu \rightarrow \nu_\tau$ reads
\begin{equation}
P (\nu_\mu \rightarrow \nu_\tau) \; \approx \; \sin^2 \left ( 1.27 ~
\frac{\Delta m^2_{23} L}{|{\bf P}|} \right ) \; ,
\end{equation}
which implies $\sin^2 2\theta_{\rm atm} \approx 1$. This ansatz can
also fit the small-angle MSW solution to the solar neutrino deficit,
if the $m_e/m_\mu$ and $m_\mu/m_\tau$ corrections to $V$ are taken into 
account.

\section{Summary and comments}
\setcounter{equation}{0}

We have given a purely phenomenological interpretation of atmospheric
and solar neutrino oscillations based on the scenarios of hierarchical and
almost degenerate neutrino masses. It is emphasized that the hierarchy 
of charged lepton and neutrino masses favors a particular
parametrization of the flavor mixing matrix, and the approximate decoupling of
solar and atmospheric neutrino oscillations nautrally appears in this
scheme. If neutrinos have nearly degenerate masses, it is shown that
interesting lepton mass matrices can be obtained starting from the
symmetries of charged lepton flavor democracy and neutrino mass
degeneracy. Two possibilities to break these symmetries, which allow
the ``maximal calculability'' for the flavor mixing matrix, have been
discussed. We find that both of them have no conflict with current
requirements of hot dark matter and neutrinoless
$\beta\beta$-decay. Another possible ansatz, in which $\sin^2
2\theta_{\rm atm} \approx 1$, has also been pointed out.

Let us end this talk by giving some brief comments on the possible problem 
of $(\sin^2 2\theta, \Delta m^2)$ 
parameter-space correlation. It is known that the flavor mixing
matrix elements are in general determined by mass ratios of neutrinos
and charged leptons. This is true, in particular, for the case of
hierarchical neutrino masses without fine tuning. Hence the
oscillation parameters $\sin^2 2\theta$ and $\Delta m^2$ are both
dependent on the neutrino masses, and they are expected to have
correlation to some extent. This correlation will lead to a further
constraint on the currently allowed ranges of $\sin^2 2\theta$ and
$\Delta m^2$, which were obtained in the naive assumption of
independence between these two parameters.

This possible correlation could also undermine the prospect to observe $CP$ 
violation in the long baseline neutrino oscillation experiments. The
reason is simply that the $CP$-violating asymmetry between 
$P(\nu_\alpha \rightarrow \nu_\beta)$ and $P(\bar{\nu}_\alpha
\rightarrow \bar{\nu}_\beta)$ is proportional not only to the 
parameter
\begin{equation}
{\cal J} \; =\; s^{~}_l c^{~}_l s_\nu c_\nu s^2 c \sin\phi \; ,
\end{equation}
but also to a product of three
oscillation terms containing $\Delta m^2_{12}$, $\Delta m^2_{23}$ and
$\Delta m^2_{31}$ \cite{FX99}. Certainly $\cal J$ is in general a function of
$m_i$, thus the correlation between $\cal J$ and three oscillating
terms could not allow ${\cal J} \sim 1\%$ to be a reasonable 
estimation. At lest for the case of hierarchical neutrino masses, we
believe that $\cal J$ has little chance to be at the percent level.

Of course, the above-mentioned correlation is model-dependent and
cannot be solved at present. Some cautious treatment is however
necessary in the future, when we have got enough knowledge about
neutrino oscillations and the dynamics of neutrino mass generation.

\vspace{0.3cm}
{\large\it Acknowledgments} ~
I would like to thank T. Morozumi for inviting and
supporting me to take part in the Hiroshima Workshop. 
I am grateful to H. Fritzsch for many useful 
discussions about neutrino masses and oscillations. 
Helpful comments from 
K. Minakata and M. Tanimoto are also acknowledged.


\end{document}